\documentclass[a4paper,12pt]{article}
\usepackage[pctex32]{graphics}
\usepackage{amssymb,amsmath}

\begin{document}

\topmargin 0pt
\oddsidemargin 0mm
\newcommand{\be}{\begin{equation}}
\newcommand{\ee}{\end{equation}}
\newcommand{\ba}{\begin{eqnarray}}
\newcommand{\ea}{\end{eqnarray}}
\newcommand{\fr}{\frac}

\renewcommand{\thefootnote}{\fnsymbol{footnote}}

\begin{titlepage}

\vspace{5mm}
\begin{center}
{\Large \bf Topologically massive gravity on AdS$_2$ spacetimes}

\vskip .6cm
 \centerline{\large Yun Soo Myung$^{1,a}$,
 Yong-Wan Kim $^{1,b}$,
and Young-Jai Park$^{2,c}$}

\vskip .6cm

{$^{1}$Institute of Basic Science and School of Computer Aided
Science,
\\Inje University, Gimhae 621-749, Korea \\}

{$^{2}$Department of Physics, BK21 Program Division, Sogang University, Seoul 121-742,
Korea}
\end{center}

\vspace{5mm}
 \centerline{{\bf{Abstract}}}

\vspace{5mm}

We study the topologically massive gravity with a negative
cosmological constant on AdS$_2$ spacetimes by making use of
Kaluza-Klein dimensional reduction. For a constant dilaton, this
two-dimensional model  admits three AdS$_2$ vacuum solutions,
which are related to AdS$_3$ and warped AdS$_3$ with an
identification upon uplifting three dimensions. We carry out
perturbation analysis around these backgrounds to find what is a
physically propagating field. It turns out that the scalar
perturbation $f$ of the Maxwell field $F_{\mu\nu}$ is
nonpropagating in the absence of gravitational Chern-Simons
terms, while it becomes a  propagating mode with mass in the
presence of gravitational Chern-Simons terms. This shows clearly
that the number of propagating degrees of freedom is one for the
topologically massive gravity. Moreover, at the points of $K=\pm
l,l/3$, the mode $f$ becomes a massless scalar which implies that there
exists still a propagating degree of freedom at the chiral points.

\vspace{5mm}


\vskip 0.8cm

\noindent $^a$ysmyung@inje.ac.kr \\
\noindent $^b$ywkim65@gmail.com \\
\noindent $^c$yjpark@sogang.ac.kr

\noindent
\end{titlepage}

\newpage

\renewcommand{\thefootnote}{\arabic{footnote}}
\setcounter{footnote}{0} \setcounter{page}{2}

\section{Introduction}
The gravitational Chern-Simons terms in three dimensional (3D)
Einstein gravity produce a physically propagating massive
graviton~\cite{DJT}. This topologically massive gravity with a
negative cosmological constant $\Lambda=-1/l^2$ (TMG$_\Lambda$)
gives us the AdS$_3$ solution~\cite{LSS1,LSS2}. For the positive
Newton's constant $G_3$,  massive modes carry negative energy on the
AdS$_3$. In this case, the AdS$_3$ is not a stable vacuum. The
opposite case of $G_3<0$ may cure the problem, but it may induce a
negative mass for the BTZ black hole.

There is a way  of  avoiding negative energy by choosing the chiral
point of $ K=l$ with the gravitational Chern-Simons coupling
constant $K$. At this point, a massive graviton becomes a massless
left-moving graviton, which carries no energy.  It may be considered
as pure gauge. However, the chiral  point raises many questions on
physical degrees of freedom
(DOF)~\cite{CDWW1,GJ,GKP,Park,GJJ,CDWW2,Carl,Stro}.

Before we proceed, it is noted that the gravitational Chern-Simons
terms are not invariant under coordinate transformations though they
are conformally invariant~\cite{GIJP,GK}. The variation of the
gravitational Chern-Simons terms plays a role to find new solutions
because it provides the  Cotten tensor term in the equation of
motion.  It is  known that the 3D Einstein gravity is locally
trivial, while its quantum status remains unclear.  All solutions to
the 3D Einstein gravity   are  also solutions to the TMG$_\Lambda$.
Furthermore, one needs to seek another method to investigate the
TMG$_\Lambda$ since it is likely  a candidate for a nontrivial 3D
gravity. One may introduce a conformal transformation to isolate the
conformal DOF (dilaton mode). Then, the Kaluza-Klein
ansatz\footnote{However, if one does not include the full tower of
Kaluza-Klein states, it does not capture the full theory.
Particularly, this is true if one confines to a reduction which  was
made by assuming an isometry. For topologically massive gravity
without a negative cosmological constant, it was known that the
exact theory has no nontrivial solutions that admits a
hypersurface-orthogonal Killing vector~\cite{AN}. In this work, the
$U(1)$ isometry used in the Kaluza-Klein reduction must not be
hypersurface-orthogonal, which means  that the vector field $A_\mu$
must be nonzero. We state clearly that the process of making a
Kaluza-Klein reduction to obtain the 2DTMG$_\Lambda$ might miss some
degrees of freedom in the original TMG$_\Lambda$.} is used to obtain
an effective two-dimensional action (2DTMG$_\Lambda$), which will be
a gauge and coordinate invariant action. Saboo and Sen~\cite{SSen}
have used the 2DTMG$_\Lambda$ to obtain the entropy of extremal BTZ
black hole by using the entropy function formalism. This is possible
because AdS$_2$ is a stable attractor solution of equations, which
governs how near-horizon geometry changes as a degenerate horizon is
being approached. On the other hand, for a constant dilaton, the
authors in~\cite{AFM} have introduced  the entropy function approach
to find three AdS$_2$ vacuum solutions of the 2DTMG$_\Lambda$:
AdS$_2$ with a positive charge, AdS$_2$ with a negative charge, and
warped AdS$_2$ with a positive charge. Upon uplifting these
solutions to three dimensions, they have obtained geometric
solutions which are either AdS$_3$ or warped AdS$_3$ with an
identification.

Hence it is very interesting to perform   perturbation analysis of
the 2DTMG$_\Lambda$, which shows  what kind of field is really
propagating on AdS$_2$ background. It is well known that there is no
physical DOF for a graviton $h_{mn}$ propagating on AdS$_3$ in 3D
Einstein gravity. Therefore, we must have zero DOF on AdS$_2$ after
Kaluza-Klein reduction of 3D Einstein gravity: a two-dimensional
graviton $h_{\mu\nu}=-h\bar{g}_{\mu\nu}$ (unphysical mode with $-1$
DOF), a gravivector $\delta F_{10}=-f$ (Maxwell mode with $0$ DOF),
and a graviscalar $\varphi$ (dilaton mode with 1 DOF).  However, we
expect to have a massive mode in the 2DTMG$_\Lambda$ because the
gravitational Chern-Simons terms could generate a massive mode in
the 3DTMG$_\Lambda$. It is supposed to be a mode like a gravivector.

In this work, we carry out perturbation analysis of the
2DTMG$_\Lambda$ around three AdS$_2$ backgrounds. We show  that  a
scalar perturbation $f$ of the Maxwell field $F_{\mu\nu}$ is trivial
and nonpropagating in the absence of the Chern-Simons terms, while
it becomes a massive propagating mode  in the presence of the
Chern-Simons terms. This shows clearly that the number of
propagating DOF is one scalar  for the 2DTMG$_\Lambda$. Moreover, at
the chiral point of $K=\pm l,~l/3$, the mode $f$ becomes a massless
scalar, which implies that there exists still a propagating DOF at
this point.

\section{Topologically massive gravity }
We start with the action for  topologically massive gravity with a
negative cosmological constant given by~\cite{DJT}
\begin{equation} \label{tmg}
I_{\rm TMG_\Lambda}=\frac{1}{16 \pi G_3}\int
d^3x\sqrt{-g}\Bigg[R_3 -2\Lambda +
\frac{K}{2}\varepsilon^{lmn}\Gamma^p_{~lq}
\Big(\partial_{m}\Gamma^q_{~np}+\frac{2}{3}\Gamma^q_{~mr}\Gamma^r_{~np}\Big)\Bigg],
\end{equation}
where $\varepsilon$ is the tensor  defined by $\epsilon/\sqrt{-g}$
with $\epsilon^{012}=1$. We choose the positive Newton's constant
$G_3>0$. The Latin indices of $l,m,n, \cdots$ denote three
dimensional tensors. The $K$-term is called the gravitational
Chern-Simons terms. Here we choose ``+" sign in the front of $K$
to avoid negative graviton energy~\cite{GJ}. It is the first
higher derivative correction in three dimensions because it is the
third-order derivative.

Varying this action leads to the Einstein equation
\begin{equation} \label{eineq}
G_{mn} + KC_{mn}=0,
\end{equation}
where the Einstein tensor including the cosmological constant is
given by
\begin{equation}
G_{mn}=R_{3mn}-\frac{R_3}{2}g_{mn}
-\frac{1}{l^2}g_{mn}
\end{equation}
and the Cotton tensor is defined by
\begin{equation}
C_{mn}= \varepsilon_m~^{pq}\nabla_p
\Big(R_{3qn}-\frac{1}{4}g_{qn}R_3\Big).
\end{equation}
We note that the Cotton tensor $C_{mn}$ vanishes for
 any solution to 3D Einstein gravity, so all solutions of general
 relativity are also solutions of the TMG$_\Lambda$. Hence, for $K=0$, the BTZ black hole~\cite{BTZ} appears as
 a  solution to
Eq. (\ref{eineq})
\begin{equation}
ds^2_{BTZ} = -N^2(r) dt^2 + \frac{dr^2}{N^2(r)} + r^2 \Big[d\theta
+ N^\theta(r) dt\Big]^2 , \label{btzmetric}
\end{equation}
where the squared lapse $N^2(r)$ and the angular shift $N^\theta(r)$
take the forms
\begin{equation}
N^2(r) = -8 G_3 m + \frac{r^2}{l^2} + \frac{16 G_3^2 j^2}{r^2},~~
N^\theta(r) = - \frac{4 G_3 j}{r^2}.\label{def_N}
\end{equation}
Here $m$ and $j$ are the mass and angular momentum of the BTZ
black hole, respectively.

On the other hand, the warped black hole solution to the
TMG$_{\Lambda}$ was first considered  in~\cite{BCl,ACGL}, and a
generalized warped black hole solution to the  TMG without a
cosmological constant was shown in ~\cite{ACL}.  For
$\nu^2=(\frac{l}{3K})^2>1$, the warped black hole solution, which is
asymptotic to warped AdS$_3$, is allowed as
\cite{StromingerWp,CD,OK}
\begin{equation}
ds^2_{wBH}=-\tilde{N}^2dt^2+\frac{l^4}{4\tilde{R}^2\tilde{N}^2}dr^2+l^2\tilde{R}^2\Big(d\theta+\tilde{N}^\theta
dt\Big)^2,
\end{equation}
where
\begin{eqnarray}
 && \tilde{R}^2(r)=\frac{r}{4}\left[3(\nu^2-1)r +(\nu^2+3)(r_++r_-)-4\nu\sqrt{r_+r_-(\nu^2+3)}\right],\nonumber\\
 && \tilde{N}^2(r)=\frac{l^2(\nu^2+3)(r-r_+)(r-r_-)}{4\tilde{R}^2(r)},\nonumber\\
 && \tilde{N}^\theta(r)=\frac{2\nu r-\sqrt{r_+r_-(\nu^2+3)}}{2\tilde{R}^2(r)}.
\end{eqnarray}
We note that for $\nu^2=1$, this metric reduces to the BTZ metric
 (\ref{btzmetric}) in a rotating frame.

 We would like to mention the AdS$_3$ solution (\ref{btzmetric}) with
 $m=-1/8G_3$ and $j=0$. It is very important to find  physical DOF  of  a massive
mode  propagating  on the AdS$_3$ background when the Chern-Simons
terms are present. There is no DOF in 3D Einstein gravity because
there is $N_c=D(D-3)/2$ DOF in $D$ dimensional Einstein gravity
\footnote{This is  obtained by counting graviton $h_{mn}$:
$D(D+1)/2-1$ (symmetric traceless tensor) $-D$ (gauge-fixing)$-D+1$
(residual gauge-fixing)=$D(D-3)/2$. Thus, one has $N_c=2, 0, -1$ for
$D=4,3,2$, respectively.  For a massive graviton, $N_c$ is changed
into $N^m_c=D(D+1)/2-D-1=(D-2)(D+1)/2$ because there is no residual
gauge symmetry.  Hence, one has $N^m_c=5,2,0$ for $D=4,3,2$,
respectively. On the other hand, one has $D-2$ DOF for a massless
vector propagation because of gauge-fixing ($-1$) and residual
gauge-fixing ($-1$), while a massive vector propagation has $D-1$
DOF. Especially, for $D=2$,  we have no DOF for a massless vector
propagation but one DOF for a massive vector propagation. }. Hence
it is evident that all components of $h_{mn}$ belong to gauge DOF.

On the other hand, we have $N^m_c=2$ DOF for a massive graviton in
3D Einstein gravity when including   the Pauli-Fierz bilinear term
of $I_{PF}= m^2_{PF}\int
d^3x\sqrt{-\bar{g}}[h^{mn}h_{mn}-(h_m~^m)^2]$~\cite{FP,Berg}
 as  a conventional mass term.
However, it is worth to mention that the bilinear Chern-Simons term
$K\int d^3x \sqrt{-\bar{g}}h^{mn}{\cal C}^{(1)}_{mn}$~\cite{LSS1}
differs from the  Pauli-Fierz term. Thus, the TMG$_\Lambda$ may
possess different physical DOF propagating on the AdS$_3$
spacetimes. Park has found $N^m_c=3$~\cite{Park}, Carlip has
obtained $N^m_c=1$~\cite{Carl}, while Grumiller et al.~\cite{GJJ}
have found $N^m_c=1$ in the chiral version of the
theory~\cite{LSS1}. Recently, it was shown that $N^m_c=1$, when
using the full power of Dirac's method for  a constrained
Hamiltonian system~\cite{BC}. Moreover, at the chiral point of
$K=l$, a massive graviton $\psi^M_{mn}$ turned out to be a
left-moving graviton $\psi^L_{mn}$ (gauge degrees of
freedom)~\cite{LSS1,ssol}. Grumiller and Johanson~\cite{GJ} have
newly introduced a physical field
$\psi^{new}_{mn}=\partial_{l/K}\psi^M_{mn}|_{K=l}$ as a logarithmic
parter of $\psi^L_{mn}$~\cite{GJ} based on the
LCFT~\cite{LCFT1,LCFT2,LCFT3,LCFT4,LCFT5,LCFT6}. However, it was
reported that $\psi^{new}_{mn}$ might not be a physical field at the
chiral point, since  it satisfies fourth-order equation and thus it
becomes a pair of dipole ghost fields with
$\psi^L_{mn}$~\cite{MyungL}.

Hence it would be better to use another method to show the existence
of a massive mode in the TMG$_{\Lambda}$. Furthermore,  to find  a
clear picture for a propagating mode at the chiral point, we study
the effective two dimensional theory by making use of Kaluza-Klein
reduction.

\section{Dimensional reduction of TMG$_\Lambda$}
We first make a conformal transformation and then perform
Kaluza-Klein dimensional reduction  by choosing  the
metric~\cite{GIJP,GK}
\begin{equation}
ds^2=\phi^2\Big[g_{\mu\nu}(x)dx^\mu dx^\nu+\Big(d\theta+A_\mu
(x)dx^\mu\Big)^2\Big].
\end{equation}
Here $\theta$ is a coordinate that parameterizes an $S^1$ with a
period $2 \pi l$. Hence, its isometry is factorized as ${\cal
G}\times U(1)$. After the ``$\theta$"-integration, the action
(\ref{tmg}) reduces to its effective two-dimensional action
(2DTMG$_\Lambda$)
\begin{eqnarray}\label{2Daction}
 I_{\rm 2DTMG_\Lambda}&=& \frac{l}{8G_3}\int d^2x \sqrt{-g}\left(\phi R+\frac{2}{\phi}g^{\mu\nu}\nabla_\mu\phi\nabla_\nu\phi
     +\frac{2}{l^2}\phi^3-\frac{1}{4}\phi F_{\mu\nu}F^{\mu\nu}\right)\nonumber\\
 &+& \frac{Kl}{32 G_3} \int d^2x \left(R\epsilon^{\mu\nu}F_{\mu\nu}+
     \epsilon^{\mu\nu}F_{\mu\rho}F^{\rho\sigma}F_{\sigma\nu}\right).
\end{eqnarray}
Here $R$ is the 2D Ricci scalar with $R_{\mu\nu}-Rg_{\mu\nu}/2=0$
and $\phi$ is a dilaton as a graviscalar. Also, the Maxwell field
$F_{\mu\nu}=2\partial_{[\mu}A_{\nu]}$ is defined by  a gravivector
$A_\mu$ and $\epsilon^{\mu\nu}$ is a tensor density. The Greek
indices of $\mu,\nu, \rho, \cdots$ represent two dimensional
tensors. Hereafter we choose $G_3=l/8$ for simplicity. It is noted
that this action was used to derive the entropy of extremal BTZ
black hole by applying the entropy function approach~\cite{SSen}.

Introducing  a dual scalar $F$ of the Maxwell field defined
by~\cite{GIJP,GK}
\begin{equation}
F=-\frac{1}{2\sqrt{-g}}\epsilon^{\mu\nu}F_{\mu\nu},
\end{equation}
equations of motion for $\phi$ and $A_\mu$ are simplified,
respectively, as
\begin{eqnarray}
 &&
  \label{EOM-phi}
 R+\frac{2}{\phi^2}(\nabla\phi)^2-\frac{4}{\phi}\nabla^2\phi+\frac{6}{l^2}\phi^2+\frac{1}{4}F^2=0,\\
 && \label{EOM-A}
  \epsilon^{\mu\nu}\partial_\nu\left[\phi
  F-\frac{K}{2}(R+3F^2)\right]=0.
\end{eqnarray}
 The equation of motion for the metric $g_{\mu\nu}$ takes
the form
\begin{eqnarray}\label{EOM-g}
 && g_{\mu\nu}\left(\nabla^2\phi-\frac{1}{l^2}\phi^3+\frac{1}{4}\phi F^2-\frac{1}{\phi}(\nabla\phi)^2\right)
    +\frac{2}{\phi}\nabla_{\mu}\phi\nabla_{\nu}\phi-\nabla_{\mu}\nabla_{\nu}\phi
   \nonumber\\
 &&
 -\frac{K}{2}\left[g_{\mu\nu}\left(\nabla^2F+F^3+\frac{1}{2}RF\right)-\nabla_{\mu}\nabla_{\nu}F\right]=0.
\end{eqnarray}
The trace part of Eq. (\ref{EOM-g})
\begin{equation}
 \label{trgEOM}
 \nabla^2\phi-\frac{2}{l^2}\phi^3+\frac{1}{2}\phi F^2
 -K\left(\frac{1}{2}RF+F^3+\frac{1}{2}\nabla^2F\right)=0
\end{equation}
 is relevant to our perturbation study.
On the other hand, the traceless part is given by
\begin{equation}
 \label{trlessgEOM}
 g_{\mu\nu}\left(\frac{1}{2}\nabla^2\phi-\frac{1}{\phi}(\nabla\phi)^2\right)
    +\frac{2}{\phi}\nabla_{\mu}\phi\nabla_{\nu}\phi-\nabla_{\mu}\nabla_{\nu}\phi
    -\frac{K}{4}g_{\mu\nu}\nabla^2F+\frac{K}{2}\nabla_{\mu}\nabla_{\nu}F=0.
\end{equation}

Now, we are in a position  to find  AdS$_2$ spacetimes as a vacuum
solution to Eqs. (\ref{EOM-phi}), (\ref{EOM-A}), and (\ref{trgEOM}).
In  case of a constant dilaton, from Eqs. (\ref{EOM-phi}) and
(\ref{trgEOM}), we have the condition of a vacuum state
\begin{equation}
(3KF-2\phi)\left(\frac{\phi^2}{l^2}-\frac{1}{4}F^2\right)=0,
\end{equation}
which provides  three distinct relations between $\phi$ and $F$
\begin{eqnarray}
\label{warp}
&& \phi_{\pm} = \pm\frac{l}{2}F, \\
\label{ads} && \phi_{w} = \frac{3K}{2}F.
\end{eqnarray}
We note that for $K=l/3$, $\phi_w=\phi_+$ which implies that it is
a degenerate vacuum.

Assuming the line element preserving ${\cal G}=SL(2,R)$ isometry
\begin{equation}\label{ads11}
ds^2_{AdS_2}=v\left(-r^2dt^2+\frac{dr^2}{r^2}\right),
\end{equation}
we have the AdS$_2$ spacetimes, which satisfy
\begin{equation} \label{ads2}
\bar{R}=-\frac{2}{v},
~~~\bar{\phi}=u,~~~\bar{F}=e/v~(\bar{F}_{10}=e),
\end{equation}
where $\bar{F}_{10}=\partial_1 \bar{A}_{0}-\partial_0 \bar{A}_{1}$ 
with $\bar{A}_{0}=er$ and $\bar{A}_{1}=0$.
In order to find the whole solution of the AdS$_2$-type, we may use
the entropy function formalism~\cite{AFM} because it provides an
efficient way of finding AdS$_2$ solution as well as the entropy of
corresponding extremal black hole.  To this end, the entropy
function is defined as
\begin{eqnarray}
{\cal E}(u,v,e,q)=2\pi\Big[qe-{\cal F}(u,v,e)\Big]
\end{eqnarray}
where ${\cal F}(u,v,e)$ is the Lagrangian density ${\cal L}_{\rm
2DTMG_\Lambda}$ evaluated when using Eq. (\ref{ads2}),
\begin{equation}
 {\cal F}(u,v,e)=-2u+\frac{2u^3v}{l^2}+\frac{ue^2}{2v}
   +\frac{K}{2}\left(\frac{2e}{v}-\frac{e^3}{v^2}\right).
\end{equation}
Upon the variation of ${\cal E}$ with respect to $u$, $v$, and
$e$, equations of motion are obtained as
\begin{eqnarray}
 \label{fequ}&& -2+\frac{6u^2v}{l^2}+\frac{e^2}{2v}=0,\\
 \label{feqv}&& \frac{2u^3}{l^2}-\frac{ue^2}{2v^2}
    -K\left(\frac{e}{v^2}-\frac{e^3}{v^3}\right)=0,\\
\label{feqe} &&
q-\frac{ue}{v}-\frac{K}{2}\left(\frac{2}{v}-\frac{3e^2}{v^2}\right)=0.
\end{eqnarray}
Equations (\ref{fequ}) and (\ref{feqv}) are those obtained by
plugging Eq. (\ref{ads2}) into Eqs. (\ref{EOM-phi}) and
(\ref{trgEOM}). This means that the entropy function formalism
uses mainly the Einstein equation and dilaton equation in the
near-horizon geometry  AdS$_2 \times S^1$ of  extremal BTZ black
hole. A difference is that Eq. (\ref{EOM-A}) is trivially
satisfied with Eq. (\ref{ads2}), while Eq. (\ref{feqe}) is working
for deriving the entropy.

Here we obtain three kinds of the AdS$_2$ solution:

\noindent(1) For $u=le/2v~(\phi_+=Fl/2)$, one has the AdS$_2$
solution with a positive charge $q$
\begin{equation}
u=\sqrt{\frac{ql^2}{2(l-K)}},~~~v=\frac{l-K}{2q},~~~e=\sqrt{\frac{l-K}{2q}}.
\end{equation}
(2) For $u=-le/2v~(\phi_-=-Fl/2)$,  the AdS$_2$ solution with a
negative charge $q$ is found as
\begin{equation}
u=-\sqrt{\frac{-ql^2}{2(l+K)}},~~~v=-\frac{l+K}{2q},~~~e=-\sqrt{\frac{l+K}{-2q}}.
\end{equation}
(3) For $u=3Ke/2v~(\phi_w=3FK/2)$, the warped AdS$_2$ solution with
a positive charge $q$ is given by
\begin{equation}
u=\sqrt{\frac{9Kql^2}{l^2+27K^2}},~~~v=\frac{K}{q},~~~e=\sqrt{\frac{4Kl^2}{q(l^2+27K^2)}}.
\end{equation}
Then, corresponding entropies of the extremal black holes are
given by
\begin{eqnarray}
 && (1)~S_+=  \frac{2\pi}{e}(l-K) \sim 2\pi\sqrt{\frac{ql}{6}\times \frac{12(l-K)}{l}},~~~l\ge K\\
 && (2)~S_-= \frac{2\pi}{e}(l+K) \sim 2\pi\sqrt{-\frac{ql}{6}\times \frac{12(l+K)}{l}},~~~l\ge -K\\
 && (3)~S_w= \frac{2\pi}{e} \frac{8Kl^2}{l^2+27K^2}\sim 2\pi\sqrt{\frac{ql}{6}\times
 \frac{96Kl}{l^2+27K^2}},~~~K>0.
\end{eqnarray}
The last relations ($\sim$) will be confirmed from the Cardy formula
if $ql$ is justified as the eigenvalue of the $L_0$-operator of dual
CFT$_2$. However, the AdS$_2$/CFT$_1$ correspondence is not still
confirmed because we have two AdS$_2$
solutions~\cite{ads21,ads22,ads23,ads24}: AdS$_2$ with a constant
dilaton and near-horizon chiral CFT$_2$ (AdS$_2$/CFT$_2$
correspondence, in this work), and AdS$_2$ with a linear dilaton and
asymptotic CFT$_1$ (AdS$_2$/CFT$_1$ correspondence). Finally, we
mention that for $K=l/3$, $S_+=\frac{4 \pi l}{3e}=S_w$ which shows a
close connection between the two solutions (1) and (3).

\section{Perturbation of 2DTMG$_{\Lambda}$ around AdS$_2$}

Now, let us consider the perturbation modes of the dilaton
(graviscalar), graviton, and dual scalar around the AdS$_2$
background as
\begin{eqnarray}
 \label{pert1} \phi&=&\bar{\phi}+\varphi, \\
 \label{pert2} g_{\mu\nu}&=&\bar{g}_{\mu\nu}+h_{\mu\nu},\\
 \label{pert3} F&=&\bar{F}(1+\delta F),
\end{eqnarray}
where the bar variables denote the AdS$_2$ background as
$\bar{\phi}=u$, $\bar{g}_{\mu\nu}=v~{\rm diag}(-r^2,r^{-2})$, and
$\bar{F}=e/v$.  This background corresponds to the near-horizon
geometry of the extremal black holes, factorized as AdS$_2 \times
S^1$. The Maxwell field has a scalar perturbation $f$ around the
background: $F_{10}=\bar{F}_{10}+\delta F_{10}$, where $\delta
F_{10}=\partial_1 a_0 -\partial_0 a_1$ and $\delta F_{10}=-f$.

In this work, we choose the particular perturbation
fields~\cite{LMK}
\begin{equation}
h_{\mu\nu}=-h\bar{g}_{\mu\nu},~~~ \delta F=\left(h-\frac{f}{e}\right).
\end{equation}
Taking the conformal gauge for $h_{\mu\nu}$  is quite reasonable for
investigating the propagation of the graviton  as an unphysical mode
on the AdS$_2$ background. Even if one considers the off-diagonal
components, these would be decoupled from $(\varphi,h,f)$. In
addition, we note that the form of $\delta F$ is determined upon
choosing the conformal gauge of $h_{\mu\nu}=-h\bar{g}_{\mu\nu}$.
Hereafter, we wish to distinguish a dual scalar perturbation
``$\delta F$" with a scalar perturbation ``$f$". The former is
useful for diagonalizing process without sources, while the latter
plays  a key mode for our perturbation study with sources. In order
to select $f$ appropriately, we need to use the source condition,
which could decouple the unphysical mode $h$ from $f$ in $\delta F$.

However, we have to mention that there may be other types of perturbations,
which lead to other linearized degrees of freedom.

\subsection{$K=0$ perturbation }

First, we study the $K=0$ case  because  this  provides a
reference case. Applying the linearizing process to
(\ref{EOM-phi}), (\ref{EOM-A}), and (\ref{trgEOM}), the perturbed
equations are given by
\begin{eqnarray}
  && \label{pEOM-phi}
 \bar{\nabla}^2h -\frac{2h}{v} -\frac{4}{u}\bar{\nabla}^2\varphi +\frac{12}{l^2}u\varphi
 +\frac{e^2}{v^2}\left(h-\frac{f}{e}\right)=0,\\
 && \label{pEOM-A}
\epsilon^{\mu\nu}\partial_\nu\Bigg[ \varphi+u\left(h-\frac{f}{e}\right)\Bigg]=0, \\
 && \label{pEOM-induced1}
 \bar{\nabla}^2\varphi-\frac{6}{l^2}u^2\varphi+\frac{ue^2}{v^2}\left(h-\frac{f}{e}\right)
 +\frac{e^2}{2v^2}\varphi=0.
\end{eqnarray}
Here we use $\delta R(h)=\delta
R_{\mu\nu}(h)\bar{g}^{\mu\nu}-\bar{R}_{\mu\nu}h^{\mu\nu}=\bar{\nabla}^\mu\bar{\nabla}^\nu
h_{\mu\nu} -\bar{\nabla}^\sigma\bar{\nabla}_\sigma
h_\mu~^\mu-\bar{R} h_\mu~^\mu/2$~\cite{LSS1}. The  perturbed
equation (\ref{pEOM-A}) for the $F$-field  leads to
\begin{equation} \label{redun}
\varphi +u \delta F=0,
\end{equation}
which implies that the dual scalar $\delta F$ is a redundant mode
(nothing but $\varphi$). For the AdS$_2$ solutions of
$u=\pm\frac{le}{2v}$, the perturbed equation (\ref{pEOM-induced1})
together with Eq. (\ref{redun}) leads to  equation for the dilaton
\begin{equation}\label{pEOM-induced12}
\left(\bar{\nabla}^2-\frac{2}{v}\right)\varphi=0,
\end{equation}
which may indicate  that the dilaton mode as a scalar is propagating on
the AdS$_2$-background~\cite{BM}. On the other hand, the perturbed
equation (\ref{pEOM-phi}) combined  with  (\ref{redun}) and
(\ref{pEOM-induced12}) leads to the fourth-order equation for the
graviton mode $h$
\begin{equation}\label{pEOM-phi1}
\left(\bar{\nabla}^2-\frac{2}{v}\right)^2h=0.
\end{equation}

It is known  from the counting of DOF  that all of these modes
belong to pure gauge because there is no physical DOF for the
graviton $h_{mn}$ propagating on the AdS$_3$ background in the
original 3D Einstein gravity. However, it seems that two modes of
$\varphi$ and $\delta F$ are propagating on the AdS$_2$ background.
Also, $h$ seems to be a propagating mode, even it satisfies the
fourth-order equation. Therefore, it is necessary to show that all
modes of $\varphi,h,f$ are nonpropagating on the AdS$_2$ background
and then apply the same method to the linearized excitations in the
$K\not=0$ TMG$_{\rm \Lambda}$ to determine which modes are
propagating. Hence we need to employ a proper method to find
physical DOF beyond the naive counting of DOF.

For this purpose, we compute the on-shell exchange amplitude
(\ref{actionK0})  in Appendix A by plugging external sources
$T,J_\varphi,J_f$ into Eqs. (\ref{pEOM-phi}), (\ref{pEOM-induced1})
and (\ref{redun}). Surprisingly, $J_{\varphi}$ completely disappears
in this amplitude. It is important to note that when choosing the
source condition of $T=\frac{e^2}{v^2}J_f$, the fourth-order
unphysical pole disappears. Simultaneously,  the second order pole
disappears under this condition. Actually, imposing this proper
source condition is equivalent to a decoupling process of the
unphysical mode $h$ from the Maxwell mode $f$ in the dual scalar
$\delta F$. In this case, the on-shell exchange amplitude
(\ref{actionK0}) reduces to that of the Maxwell mode $f$ solely
\begin{equation}
\label{reducedK0}
   \bar{A}^{K=0}=\frac{1}{2uv}\int d^2p
      [J^2_f].
\end{equation}
As a result, the effective 2D gravity theory, which is matched
with the original 3D Einstein gravity, has no physically
propagating modes. Based on this analysis, we  explore the effect of
the gravitational  Chern-Simons terms in the next section.

\subsection{$K\neq0$ perturbation }

For $K\neq0$, the perturbed equations of motion  are complicated
to have
\begin{eqnarray}
 && \label{pEOMK-phi}
 \bar{\nabla}^2h-\frac{2}{v}h-\frac{4}{u}\bar{\nabla}^2\varphi+\frac{12}{l^2}u\varphi
 +\frac{e^2}{v^2}\delta F=0,\\
 && \label{pEOMK-A}
 \epsilon^{\mu\nu}\partial_\nu\Bigg[\frac{e}{v}\Big(\varphi+u\delta
 F\Big)
 -\frac{K}{2}\Big(\bar{\nabla}^2h-\frac{2}{v}h+\frac{6e^2}{v^2}\delta F\Big)\Bigg]=0, \\
 && \label{pEOMK-g1}
 \bar{\nabla}^2\varphi-\frac{6}{l^2}u^2\varphi+\frac{e^2}{2v^2}\varphi+\frac{ue^2}{v^2}\delta F\nonumber\\
 &&
 -K\Bigg(\frac{e}{2v}\bar{\nabla}^2h-\frac{e}{v^2}h+\frac{e(3e^2-v)}{v^3}\delta F
 +\frac{e}{2v}\bar{\nabla}^2\delta F\Bigg)=0.
\end{eqnarray}
Eq. (\ref{pEOMK-A}) implies
\begin{equation} \label{pEOMK-AA}
\frac{e}{v}\Big(\varphi+u\delta F\Big)
 -\frac{K}{2}\Big(\bar{\nabla}^2h-\frac{2}{v}h+\frac{6e^2}{v^2}\delta
 F\Big)=0.
 \end{equation}
Solving Eq. (\ref{pEOMK-AA}) for  $\delta F$ and inserting it into
Eq. (\ref{pEOMK-g1}), we obtain
\begin{equation}
 \bar{\nabla}^2\varphi-\left(\frac{6u^2}{l^2}+\frac{e^2}{2v^2}\right)\varphi
 -\frac{Ke}{2v}\left(\bar{\nabla}^2-\frac{2}{v}\right)\delta F=0.
\end{equation}
Since the second term can be further simplified as $(2/v)\varphi$,
irrespective of  the AdS$_2$ solutions (1), (2) and the warped
AdS$_2$ solution (3), we find the same diagonalized equation as
\begin{equation}
\label{pEOMK-g2}
 \left(\bar{\nabla}^2-\frac{2}{v}\right)
 \left(\varphi-\frac{Ke}{2v}\delta F\right)=0
\end{equation}
which  may correspond to Eq. (\ref{pEOM-induced12}) for $K=0$.
Moreover,  solving Eq. (\ref{pEOMK-AA}) for
$(\bar{\nabla}^2-\frac{2}{v})h$ and inserting it into Eq.
(\ref{pEOMK-phi}), we obtain a coupled equation
\begin{equation}
 \label{pEOMK-phi1}
 \bar{\nabla}^2\varphi
 -\left(\frac{ue}{2vK}+\frac{3u^2}{l^2}\right)\varphi
 -\left(\frac{u^2e}{2vK}-\frac{5e^2u}{4v^2}\right)\delta F=0.
\end{equation}
It reduces to
\begin{equation}
 \label{pEOMK-phi2}
 \left(\bar{\nabla}^2-\frac{2}{v}\right)\varphi
 -\left(\frac{ue}{2vK}-\frac{5}{4v}\right)\Big(\varphi+u\delta F\Big)=0
\end{equation}
for the AdS$_2$ solutions (1) and (2), while it takes the form
\begin{equation}
 \label{pEOMK-phi3}
 \left(\bar{\nabla}^2-\frac{2}{v}\right)\varphi
 -\frac{1}{v}\left(\frac{l^2-27K^2}{l^2+27K^2}\right)\varphi
 +\frac{2l^2u}{v(l^2+27K^2)}\delta F=0
\end{equation}
for the warped AdS$_2$ solution (3).

Making use of Eq. (\ref{pEOMK-g2}),  Eqs. (\ref{pEOMK-phi2}) and
(\ref{pEOMK-phi3}) could be  rewritten as two coupled equations
for $\varphi$ and $\delta F$
\begin{equation}
 \label{pEOMK-phi4}
 \left(\bar{\nabla}^2-\frac{2}{v}\right)\delta F
 -\frac{2v}{Ke}\left(\frac{ue}{2vK}-\frac{5}{4v}\right)\Big(\varphi+u\delta F\Big)=0,
\end{equation}
\begin{equation}
 \label{pEOMK-phi5}
 \left(\bar{\nabla}^2-\frac{2}{v}\right)\delta F
 -\frac{2}{Ke}\left(\frac{l^2-27K^2}{l^2+27K^2}\right)\varphi
 +\frac{2u}{Ke}\left(\frac{2l^2}{l^2+27K^2}\right)\delta F=0.
\end{equation}
Acting $\left(\bar{\nabla}^2-\frac{2}{v}\right)$ on Eq.
(\ref{pEOMK-phi4}), and then eliminating $\varphi$ again by using
Eq. (\ref{pEOMK-g2}), we arrive at the diagonalized equation for the
dual scalar $\delta F$
\begin{equation}
 \label{pEOMK-phi6}
 \left(\bar{\nabla}^2-\frac{2}{v}\right)\left(\bar{\nabla}^2-m^2_{\pm}\right)
 \delta F=0,
\end{equation}
for the AdS$_2$ solutions (1) and  (2). Here, the mass $m^2_{\pm}$
is given by
\begin{equation}
 \label{masspm}
 m^2_{\pm}=\frac{2}{v}-\frac{1}{4v}\left(1\pm\frac{l}{K}\right)\left(5\mp\frac{l}{K}\right),
\end{equation}
where the upper and lower signs denote the  AdS$_2$ solutions (1)
and (2), respectively.

 For the warped AdS$_2$ solution
(3), the diagonalized equation is found from Eq.
(\ref{pEOMK-phi5}) as
\begin{equation}
 \label{pEOMK-phi7}
 \left(\bar{\nabla}^2-\frac{2}{v}\right)\left(\bar{\nabla}^2-m^2_{w}\right)
 \delta F=0
\end{equation}
whose mass  $m^2_{w}$ takes the form
\begin{equation}
\label{massw}
 m^2_{w}=-\frac{3(l^2-9K^2)}{v(l^2+27K^2)}.
\end{equation}
In deriving Eqs. (\ref{pEOMK-phi6}) and (\ref{pEOMK-phi7}),
we assume that $(\bar{\nabla}^2-2/v)\delta F\not=0$. This is because if
$(\bar{\nabla}^2-2/v)\delta F=0$, it may reduce to the $K=0$ case,
as is shown in Eq. (\ref{pEOM-induced12}) for $\varphi=-u\delta
F$. This implies that $m^2_{\pm,w}=2/v$ may be excluded in favor
of  a massive propagating mode. However, it is not true that
solutions to the fourth-order equation
$(\bar{\nabla}^2-2/v)^2\delta F=0$ are the same as those of the
second-order equation $(\bar{\nabla}^2-2/v)\delta F=0$. This may
be  similar to the behavior of scalar fields at the
Breitenlohner-Freedman bound in the AdS$_4$ spacetimes, where  the
disappearance of one branch solution is regarded as an illusion.

For $m^2_{\pm,w}\not=2/v$, Eqs. (\ref{pEOMK-phi6}) and
(\ref{pEOMK-phi7}) may imply the  second-order equations
\begin{equation} \label{sedeqs}
\left(\bar{\nabla}^2-m^2_{\pm}\right)
 \delta F=0,~~\left(\bar{\nabla}^2-m^2_{w}\right)
 \delta F=0
 \end{equation}
which seem to be more attractive than the fourth-order equations.

However, we expect to obtain  a single massive mode like $f$ which
is propagating on the AdS$_2$ background because it turned out that
$N^m_c=1$ for the $K\not=0$ case. Therefore, the second-order
equations (\ref{sedeqs}) are still far from our goal of finding $f$,
even though the explicit masses  are found. Inspired by the $K=0$
case, we wish to compute the on-shell exchange amplitude for the
$K\not=0$ case, too. Actually, this was done in  Appendix B by
plugging the external sources $T,J_\varphi,J_f$ into the linearized
equations (\ref{pEOMK-phi}), (\ref{pEOMK-g1}), and (\ref{pEOMK-AA}).
As is shown in Eq. (\ref{nonkamp}), its form is very complicated,
including  poles of
\begin{equation}
\frac{1}{p^2+m^2},~~\frac{1}{(p^2+\frac{2}{v})(p^2+m^2)},~~\frac{1}{(p^2+\frac{2}{v})^2(p^2+m^2)}
\end{equation}
with the mass $m^2=m^2_{\pm,w}$. We note that Eqs.
(\ref{pEOMK-phi6}) and (\ref{pEOMK-phi7}) correspond to the second
pole in the above. However, it is very difficult to isolate a
physical amplitude including $\frac{1}{p^2+m^2}$ only. Hopefully,
inspired by the $K=0$ case, we may choose the same source condition
$T=\frac{e^2}{v^2}J_f$ to have a physical amplitude. Requiring  this
condition, the Fourier-transformed on-shell amplitude is reduced to
a very simple expression
\begin{equation} \label{ssexp}
 \bar{A}^K_{\pm,w}=\frac{u}{8}\int d^2p \frac{1}{{p}^2+m_{\pm,w}^2}\left(\frac{2e}{Kv}J_f-J_\varphi\right)^2,
\end{equation}
which contains $f$ and $\varphi$-amplitudes only. This confirms that
the source condition decouples the unphysical mode $h$ from
$\bar{A}^K_{\pm,w}$ effectively, remaining the $J_f$ and
$J_\varphi$-amplitudes\footnote{If one chooses the proper source
condition, Eq. (\ref{eomS1}) in Appendix B implies that $\varphi
\sim f $ even for $K\not=0$ case. This may explain partly why the
dilaton mode survives for the $K\not=0$ amplitude.}. We can  check
easily that this expression approaches the $K=0$ amplitude in the
limit of $K\rightarrow 0$: Explicitly, the action (\ref{actionK0})
is exactly recovered from (\ref{ssexp}).

It seems appropriate to comment on why the dilaton mode appears in
the $K\not=0$ amplitude, in comparison with the absence of
$J_\varphi$ in the $K=0$ amplitude. The dilaton is the conformal
mode, which is surely the pure gauge in view of the 3D Einstein
gravity. As was shown in footnote 3, we have a relation of $\phi
\sim f$ under the source condition. Hence, we may regard the
dilation as a redundant mode, even it takes the same massive pole as
the mode $f$ does.

Now we are in a position to analyze what these amplitudes are
implied.

 (1) For the AdS$_2$ solution with a positive charge, the
allowable  range is defined to be $l\ge K$. This is required for a
positive entropy.  Figure 1 shows the graph  for mass $m^2_+$ and
entropy $S_+$ versus $K$. The one-shell exchange amplitude is
given by
\begin{equation} \label{ssexp+}
 \bar{A}^K_{+}=\frac{u}{8}\int d^2p
 \frac{1}{{p}^2+m_{+}^2}\left(\frac{l}{uvK}J_f-J_\varphi\right)^2.
\end{equation}
At the chiral point of  $K=l$, we have zero mass $m^2_+=0$. Then,
its amplitude leads to
\begin{equation} \label{massless}
\bar{A}^{K=l}_{+}=\frac{u}{8}\int d^2p
\frac{1}{{p}^2}\left(\frac{J_f}{uv}-J_\varphi\right)^2,
\end{equation}
which implies that the mode $f$ is massless. Actually, this
interpretation is consistent with $N_c=1$ ~\cite{GJJ,Carl,BC},
which states that there is still a physical DOF at the chiral
point.  For $K=l/3$, we have zero mass $m^2_+=0$  but its
amplitude takes a slightly different form
\begin{equation}
\bar{A}^{K=l/3}_{+}=\frac{u}{8}\int d^2p
\frac{1}{{p}^2}\left(\frac{3J_f}{uv}-J_\varphi\right)^2.
\end{equation}
Comparing it with Eq. (\ref{massless}), this case could be also
considered as the massless  propagation at another chiral point.
\begin{figure}[t!]
   \centering
   \includegraphics{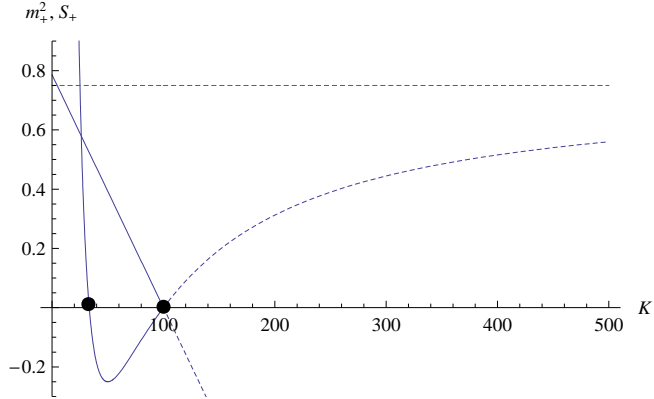}
\caption{Mass $m^2_+$ and entropy $S_+$ for the AdS$_2$ solution
with a positive charge for $l=100$, $v=1$: The curve is for the mass
and the straight line for the entropy which is scaled to compare
with the mass. The right-hand point ($\bullet$) is the chiral point
($K=l$), while the left-hand point ($\bullet$) is another chiral
point $K=l/3$. Since the AdS$_2$ solution with a positive charge is
valid for $l \ge K$, the permitted region is $0\le K\le 100$ but not
$K>l$.  } \label{fig.1}
\end{figure}

\begin{figure}[t!]
   \centering
   \includegraphics{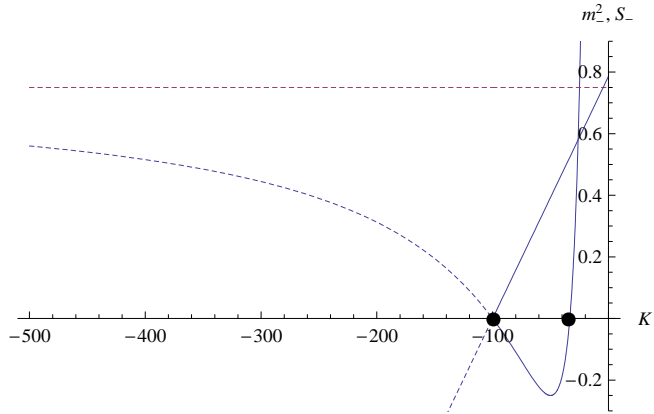}
\caption{Mass $m^2_-$ and entropy $S_-$ for the AdS$_2$ solution
with a negative charge for $l=100$, $v=1$: The curve is for the mass
and the straight line for the entropy. The right-hand point
($\bullet$) is another chiral  point ($K=-l/3$), while the left-hand
point ($\bullet$) is the chiral point ($K=-l$). Since the AdS$_2$
solution with a negative  charge is valid for $l\ge -K$, the
permitted region is $-100\le K\le 0$ but not $K<-l$ . }
\label{fig.2}
\end{figure}

\begin{figure}[t!]
   \centering
   \includegraphics{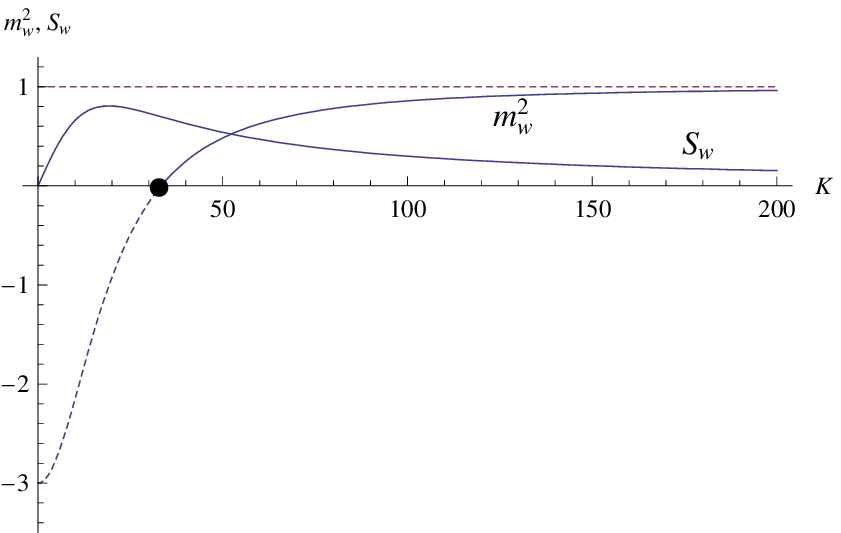}
\caption{Mass $m^2_w$ and entropy $S_w$ for the warped AdS$_2$
solution with $K \ge 0$. The point ($\bullet$) denotes a chiral
point ($K=l/3, \nu=1$). We have two limits: $m^2_w \to
-1/3$ as $K\to 0$ and $m^2_w \to 1$ as $K\to \infty$.}
\label{fig.3}
\end{figure}

(2) For the AdS$_2$ solution with a negative charge, the permitted
range is defined to be $l\ge - K$. This is required for a positive
entropy. Figure 2 shows the graph  for mass $m^2_-$ and entropy
$S_-$ versus $K$. The one-shell exchange amplitude is given by
\begin{equation} \label{ssexp-}
 \bar{A}^K_{-}=\frac{u}{8}\int d^2p
 \frac{1}{{p}^2+m_{-}^2}\left(\frac{l}{uvK}J_f+J_\varphi\right)^2.
\end{equation}
At the chiral point of  $K=-l$, we have zero mass $m^2_-=0$. Then,
its amplitude leads to
\begin{equation} \label{massless-}
\bar{A}^{K=-l}_{-}=\frac{u}{8}\int d^2p
\frac{1}{{p}^2}\left(\frac{J_f}{uv}+J_\varphi\right)^2,
\end{equation}
which implies that the mode $f$ is massless. This interpretation
is consistent with $N_c=1$ ~\cite{GJJ,Carl,BC},
which states that there is still a physical DOF at the chiral
point. For $K=-l/3$, we have zero mass $m^2_-=0$  but its
amplitude takes a different form
\begin{equation}
\bar{A}^{K=-l/3}_{-}=\frac{u}{8}\int d^2p
\frac{1}{{p}^2}\left(\frac{3J_f}{uv}+J_\varphi\right)^2.
\end{equation}
Comparing it with Eq. (\ref{massless-}), this case may be considered
as the massless propagation at another chiral point.

(3) For the warped AdS$_2$ solution, there is no restriction on
$K$ because the positive entropy is guaranteed for $K\ge 0$. The
one-shell exchange amplitude is given by
\begin{equation} \label{ssexpw}
 \bar{A}^K_{w}=\frac{u}{8}\int d^2p
 \frac{1}{{p}^2+m_{w}^2}\left(\frac{3}{uv}\frac{4l^2}{l^2+27K^2}J_f-J_\varphi\right)^2.
\end{equation}
As is shown in Fig. 3, we have zero mass $m^2_{w}=0$ for $K=l/3$.
Its amplitude  takes the form
\begin{equation}
\bar{A}^{K=l/3}_{w}=\frac{u}{8}\int d^2p
\frac{1}{{p}^2}\left(\frac{3J_f}{uv}-J_\varphi\right)^2=\bar{A}^{K=l/3}_{+}.
\end{equation}
We prove a close connection that  for $K=\frac{l}{3}$,   the
warped AdS$_2$
 amplitude $\bar{A}^K_w$ is equal to  $\bar{A}^K_+$ of the AdS$_2$ solution with
a positive charge.  For $K<l/3(\nu<1)$, $f$ has a negative mass,
while for $K>l/3(\nu>1)$, $f$ has a positive mass. It is consistent
with the propagation of the massive graviton on the warped AdS$_2$
background in the TMG$_{\rm \Lambda}$.

\section{Discussions}
We have first discussed the perturbation of the effective 2D
Einstein gravity around the AdS$_2$ background. For the $K=0$ case,
it seems that a dual scalar perturbation $\delta F=(h-f/e)$ of the
Maxwell field is equal to the dilaton mode $ \varphi$, and they
satisfy the second-order equation (\ref{pEOM-induced12}). This may
imply that the modes $\delta F$ and $\varphi$ are propagating on the
AdS$_2$ background. We note that it would be better to use the  dual
scalar $\delta F$ in the perturbation calculation, while the Maxwell
scalar $f$ is a candidate for the physical mode. Consequently, we
have successfully shown that there is no propagating modes on the
AdS$_2$ background by calculating the on-shell exchange amplitude
with the source condition of $T=\frac{e^2}{v^2} J_f$,
which arises from the diffeomorphism (gauge symmetry). In this
case, the dilaton mode $\varphi$ disappears in the amplitude. This
reflects that the effective 2D gravity, which is matched with the 3D
Einstein gravity, has no physically propagating degrees of freedom
on the AdS$_2$ background because all modes of $h,\varphi,f$ are
pure gauge.

For the $K\not=0$ case,  it seems that $\delta F$ becomes the
massive mode, which is propagating on the AdS$_2$ background.
Considering the relation (\ref{pEOMK-g2}), it is conjectured that
the dilaton mode $\varphi$ is propagating on the AdS$_2$ background,
too. However, this is not true because the unphysical graviton mode
$h$ is still involved in $\delta F$. In order to obtain physically
propagating modes, we have computed the on-shell exchange amplitude.
It takes a complicated form so that we could not isolate the
physical amplitude from the full amplitude. Fortunately, inspired by
the $K=0$ case, a proper choice of the source condition simplifies
the on-shell amplitude significantly, leaving the massive pole for
any AdS$_2$ background. As a result, we have clearly shown that the
Maxwell scalar $f$ is a truly massive mode propagating on the
AdS$_2$ background.

\section*{Acknowledgement}
The authors thank H. W. Lee for helpful discussions. Y. S. Myung  was supported by the Korea Research Foundation (KRF-2006-311-C00249)
funded by the Korea Government (MOEHRD). Y.-W. Kim was supported by the Korea
Research Foundation Grant funded by Korea Government (MOEHRD):
KRF-2007-359-C00007. Y.-J. Park was supported by
the Korea Science and Engineering Foundation (KOSEF) grant
funded by the Korea government (MOST) (R01-2007-000-20062-0).

\section*{Appendix: On-shell exchange amplitudes with external sources}

\subsection*{A. $K=0$ case}

In order to find  physically propagating modes of the effective 2D
gravity on the AdS$_2$ background, we start with the action
(\ref{2Daction}) with $K=0$. Its bilinear form is coupled with
external sources as
\begin{equation}
 \label{Saction}
 A^{K=0}=\int d^2 x \left[\delta_2 {\cal L}_{\rm 2Dgravity}(h,\varphi,f)
 +h_{\mu\nu}T^{\mu\nu}+\varphi J_\varphi+\frac{e}{v^2}fJ_f\right].
\end{equation}
Here, $\delta_2 {\cal L}_{\rm 2Dgravity}$ is the bilinear action of
effective 2D gravity around the AdS$_2$ background, and the external
sources are given by $T^{\mu\nu}(x), J_\varphi(x), J_f(x)$ for the
graviton, dilaton, and  scalar of the Maxwell field, respectively.
In order to make  the connection with Sec. 4.1, the above action
could be  rewritten in terms of the dual scalar perturbation $\delta
F$
\begin{equation}
\label{Saction1}
 A^{K=0}=\int d^2 x \left[\delta_2 {\cal L}_{\rm 2Dgravity}(h,\varphi,\delta F)
 +h(-T+\frac{e^2}{v^2}J_f)+\varphi J_\varphi-\frac{e^2}{v^2}\delta F
 J_f\right]
\end{equation}
with the trace $T=\bar{g}^{\mu\nu}T_{\mu\nu}$. In order to
investigate  what physical excitations there are, we have to
derive the linearized equations. 
At this stage, let us introduce the diffeomorphism generated by the coordinate transformation~\cite{GIJP}
 \begin{equation}
 x^\mu \to \tilde{x}^\mu=x^\mu -\epsilon^\mu(t,r).
 \end{equation}
 Then, the dilaton, gauge field, and metric transform as follows
 \begin{eqnarray}
 \delta \tilde{\varphi} &\to& \delta \varphi +\epsilon^\mu \partial_\mu \bar{\phi},  \nonumber \\
 \label{gaug1}\delta \tilde{a}_\mu &\to& \delta a_\mu +\epsilon^\nu \partial_\nu \bar{A}_\mu+\bar{A}_\nu\partial_\mu\epsilon^\nu +\partial_\mu \Lambda, \\
 \delta \tilde{h}_{\mu\nu} &\to& \delta h_{\mu\nu} +\epsilon^\rho \partial_\rho \bar{g}_{\mu\nu}+
    \bar{g}_{\mu\sigma}\partial_\nu \epsilon^{\sigma}+ \bar{g}_{\nu\sigma}\partial_\mu \epsilon^{\sigma},  \nonumber
 \end{eqnarray}
 where $a_\mu$ also undergoes an Abelian gauge transformation with gauge function $\Lambda(t,r)$.
 For the AdS$_2$ background in Eqs. (\ref{ads11}) and (\ref{ads2}), the relevant transformations lead to
 \begin{eqnarray}
        \delta \tilde{\varphi} &\to& \delta \varphi, \nonumber \\
 \label{gaug2}\delta \tilde{a}_0 &\to& \delta a_0 +e\epsilon^1 +er\partial_0\epsilon^0 +\partial_0 \Lambda,  \\
 \delta \tilde{a}_1 &\to& \delta a_1  +er\partial_1\epsilon^0 +\partial_1 \Lambda,  \nonumber\\
 \delta \tilde{h} &\to& \delta h- \partial_\rho \epsilon^\rho.  \nonumber
 \end{eqnarray}
 From these, we obtain two gauge-invariant scalars
 \begin{equation}
 \varphi, ~~\delta F=h-\frac{f}{e},
 \end{equation}
 while the scalar perturbation $f$ transforms as
 \begin{equation}
 \delta \tilde{f} \to \delta f -e \partial_\rho \epsilon^\rho.
 \end{equation}
 In general, the bilinear action $ \delta_2 {\cal L}_{\rm 2Dgravity}(h,\varphi,\delta F)$ is invariant
 under the diffeomorphisms of Eq. (\ref{gaug1}). Also this invariance should persist in when coupling with the external sources.
 Hence, considering two gauge-invariant scalars of $\varphi$ and $\delta F$ and gauge-dependent trace $ h $, the action in Eq. (\ref{Saction1}) is not invariant unless
 \begin{equation}
 T=\frac{e^2}{v^2} J_f.
 \end{equation}
 This is the origin of the source condition, which states that the AdS$_2$-background symmetry reflects
 the source-conservation laws~\cite{SSRSS}.
 {\it Hence, we have to choose this source condition whenever identifying physical
 degrees of freedom.}
 
From the action (\ref{Saction1}), we
obtain the equations of motion
\begin{eqnarray}
 && \label{EOM-As0} \varphi+u\delta F=-J_f,\\
 && \label{EOM-gs0} \bar{\nabla}^2\varphi-\frac{6}{l^2}u^2\varphi
       +\frac{e^2}{2v^2}\varphi+\frac{ue^2}{v^2}\delta F=-T, \\
 && \label{EOM-phis0} \left(\bar{\nabla}^2-\frac{2}{v}\right)h-\frac{4}{u}\bar{\nabla}^2\varphi
       + \frac{12}{l^2}u\varphi+\frac{e^2}{v^2}\delta F=-J_\varphi.
\end{eqnarray}
If the external sources are  turned off, these  are the same
equations of (\ref{redun}), (\ref{pEOM-induced1}), and
(\ref{pEOM-phi}). Hence we could follow the diagonalizing process in
Sec. 4.1 with the sources. Eliminating $\delta F$ in Eq.
(\ref{EOM-gs0}) by using (\ref{EOM-As0}), we obtain Eq.
(\ref{pEOM-induced12}) with the sources
\begin{equation}
\label{phis0}
\left(\bar{\nabla}^2-\frac{2}{v}\right)\varphi=-T+\frac{e^2}{v^2}J_f,
\end{equation}
and from Eq. (\ref{EOM-phis0}), we have the equation of motion for
$h$
\begin{equation}
\label{hs0}
 \left(\bar{\nabla}^2-\frac{2}{v}\right)h=-J_\varphi-\frac{4}{u}T+\frac{5}{uv}J_f
 -\frac{6}{uv}\varphi.
\end{equation}

To obtain the on-shell exchange amplitude induced by the sources,
we  substitute the linearized  equations (\ref{EOM-As0}),
(\ref{EOM-gs0}), and (\ref{EOM-phis0}) into
Eq.~(\ref{Saction1})~\cite{SSRSS}. Then we find the
Fourier-transformed on-shell amplitude as
\begin{equation}
\label{Saction2}
 \bar{A}^{K=0}=\frac{1}{2}\int d^2 p \left[h(p) \left(-T+\frac{e^2}{v^2}J_f \right)+\varphi(p) J_\varphi-\frac{e^2}{v^2}\delta F(p)
 J_f\right].
\end{equation}
On the other hand, from Eqs.(\ref{phis0}), (\ref{EOM-As0}) and
(\ref{hs0}), the Fourier-transformed fluctuations  are given by
\begin{eqnarray}
 \label{p11} && \varphi(p)=\frac{T-\frac{e^2}{v^2}J_f}{{p}^2+\frac{2}{v}}, \\
 \label{p12}&& \delta F(p)=-\frac{T-\frac{e^2}{v^2}J_f}{u({p}^2+\frac{2}{v})}-\frac{1}{u}J_f,\\
 \label{p13}&&  h(p)=\frac{J_\varphi+\frac{4}{u}T-\frac{5}{uv}J_f}{{p}^2+\frac{2}{v}}
         -\frac{6(T-\frac{e^2}{v^2}J_f)}{uv({p}^2+\frac{2}{v})^2}.
\end{eqnarray}
Finally, plugging Eqs. (\ref{p11})-(\ref{p13}) into
(\ref{Saction2}) leads to  the Fourier-transformed on-shell
amplitude
\begin{equation}
\label{actionK0}
   \bar{A}^{K=0}=\frac{1}{2}\int d^2p
      \left[\frac{1}{uv}J^2_f-\frac{2(T-\frac{e^2}{v^2}J_f)(2T-\frac{3e^2}{v^2}J_f)}{u({p}^2+\frac{2}{v})}
        +\frac{6(T-\frac{e^2}{v^2}J_f)^2}{uv({p}^2+\frac{2}{v})^2} \right].
\end{equation}
We note that $J_\varphi$ disappears in $\bar{A}^{K=0}$, which
ensures that $\varphi$ is  decoupled from the effective 2D gravity
theory.  It is very important to note that under the source
condition of $T=\frac{e^2}{v^2}J_f$, the fourth-order unphysical
pole (the last term in (\ref{actionK0})) vanishes.
Simultaneously, the second term disappears.
As a result, the effective 2D gravity theory, which is
matched  with the original 3D Einstein gravity, has no physical
propagation modes under the proper source condition. Based on this
method, we explore the effect of the gravitational  Chern-Simons
term in the next section.

\subsection*{B. $K\neq 0$ case}

Having the bilinear form $\delta_2{\cal L}_{\rm
2DTMG_\Lambda}(h,\varphi,f)$ of Eq. (\ref{2Daction}) instead of
$\delta_2{\cal L}_{\rm 2Dgravity}(h,\varphi,f)$ in Eq.
(\ref{Saction}), the linearized equations with the external sources
are given by
\begin{eqnarray}
 && \label{pEOMK-As}
 \frac{e}{v}\Big(\varphi+u\delta  F\Big)
 -\frac{K}{2}\Big(\bar{\nabla}^2h-\frac{2}{v}h+\frac{6e^2}{v^2}\delta F\Big)=-\frac{e}{v}J_f, \\
 && \label{pEOMK-g1s}
 \bar{\nabla}^2\varphi-\frac{6}{l^2}u^2\varphi+\frac{e^2}{2v^2}\varphi+\frac{ue^2}{v^2}\delta F\nonumber\\
 &&
 -K\Bigg(\frac{e}{2v}\bar{\nabla}^2h-\frac{e}{v^2}h+\frac{e(3e^2-v)}{v^3}\delta F
 +\frac{e}{2v}\bar{\nabla}^2\delta F\Bigg)=-T, \\
  && \label{pEOMK-phis}
 \bar{\nabla}^2h-\frac{2}{v}h-\frac{4}{u}\bar{\nabla}^2\varphi+\frac{12}{l^2}u\varphi
 +\frac{e^2}{v^2}\delta F=-J_\varphi.
\end{eqnarray}
If the above sources are  turned off, these  are the same equations
of (\ref{pEOMK-AA}), (\ref{pEOMK-g1}), and (\ref{pEOMK-phi}),
respectively. Hence we could follow the diagonalizing process in
Sec. 4.2 with the sources. Instead, we choose  an integrated
diagonalization approach to report a compact expression. By solving
Eq. (\ref{pEOMK-As}) by $\delta F$ and inserting it into the fourth
term in Eq. (\ref{pEOMK-g1s}), we have Eq. (\ref{pEOMK-g2}) with the
sources
\begin{equation}
 \label{eomS1}
 \left(\bar{\nabla}^2-\frac{2}{v}\right)\Big(\varphi- \frac{Ke}{2v}\delta F\Big)=
 -T+\frac{e^2}{v^2}J_f.
\end{equation}
We also eliminate $\left(\bar{\nabla}^2-\frac{2}{v}\right)h$ in Eq.
(\ref{pEOMK-phis}) by using Eq. (\ref{pEOMK-As}), and then apply the
operator $\left(\bar{\nabla}^2-\frac{2}{v}\right)$ to both sides.
Using  Eq. (\ref{eomS1}) leads  to Eqs. (\ref{pEOMK-phi6}) and
(\ref{pEOMK-phi7}) with the sources
\begin{eqnarray}
 \label{eomS2}
 \left(\bar{\nabla}^2-\frac{2}{v}\right)\left(\bar{\nabla}^2-m^2\right)\delta F
 &=&\left(\bar{\nabla}^2-\frac{2}{v}\right)\left[\frac{2v}{Ke}T+\left(\frac{u}{K^2}-\frac{2e}{Kv}\right)
 -\frac{uv}{2Ke}J_\varphi\right]\nonumber\\
 &-&\left(\frac{u}{K^2}-\frac{4}{Ke}+\frac{6u^2v}{Kel^2}\right)\left(T-\frac{e^2}{v^2}J_f\right),
\end{eqnarray}
where $m^2$ is defined as
\begin{equation}
m^2=-\frac{2ue}{Kv}-\frac{u^2}{l^2}\left(3+\frac{l^2}{K^2}\right).
\end{equation}
We note that $m^2$  is nothing but the masses $m^2_\pm$ for the
AdS$_2$ in Eq. (\ref{masspm}) and $m^2_w$ in Eq. (\ref{massw}) for
the warped AdS$_2$ solutions.  In addition, making use of Eq.
(\ref{eomS1}), the graviton mode $h$ satisfies
\begin{equation}
 \label{eomS3}
 \left(\bar{\nabla}^2-\frac{2}{v}\right)h=-J_\varphi-\frac{4}{u}\left(T-\frac{e^2}{v^2}J_f\right)
 +\left(\frac{8}{uv}-\frac{12u}{l^2}\right)\varphi
 +\left[\frac{2Ke}{uv}\left(\bar{\nabla}^2-\frac{2}{v}\right)-\frac{e^2}{v^2}\right]\delta  F.
\end{equation}

Similar to the $K=0$ case, after obtaining the Fourier-transformed fluctuations
from Eqs. (\ref{eomS1}), (\ref{eomS2}), (\ref{eomS3}) and making a tedious calculation,
we arrive at the Fourier-transformed on-shell amplitude induced by the external sources as
\begin{equation} \label{nonkamp}
 \bar{A}^K=\frac{1}{2}\int d^2 p
 \left[C_{TT}TT+C_{T\varphi}TJ_\varphi+C_{Tf}TJ_f++C_{\varphi\varphi}J_\varphi J_\varphi
       +C_{\varphi f}J_\varphi J_f+C_{ff}J_f J_f\right].
\end{equation}
Here  their coefficients including poles  are given by
\begin{eqnarray}
 && C_{TT}=\frac{-\frac{1}{uv}\left(\frac{4u^2v}{K^2}-\frac{12ue}{K}\right)\left({p}^2+\frac{2}{v}\right)+A}
       {\left({p}^2+\frac{2}{v}\right)^2\left({p}^2+m^2\right)},\nonumber\\
 && C_{T\varphi}=\frac{-2\left({p}^2+\frac{2}{v}\right)-\frac{Ke}{v}\left(\frac{u}{K^2}-\frac{4}{Ke}+\frac{6u^2v}{Kel^2}\right)}
       {\left({p}^2+\frac{2}{v}\right)\left({p}^2+m^2\right)},\nonumber\\
 && C_{Tf}=\frac{\frac{4e}{Kv}\left({p}^2+\frac{2}{v}\right)^2
       +\left[\frac{8e^2}{v^2}\left(\frac{u}{K^2}-\frac{3e}{Kv}\right)
       +\frac{2e^2}{v^2}\left(\frac{u}{K^2}-\frac{4}{Ke}+\frac{6u^2v}{Kel^2}\right)\right]
       \left({p}^2+\frac{2}{v}\right)-\frac{2e^2}{v^2}A}
       {\left({p}^2+\frac{2}{v}\right)^2\left({p}^2+m^2\right)},\nonumber\\
 && C_{\varphi\varphi}=\frac{u}{4\left({p}^2+m^2\right)},\nonumber\\
 && C_{\varphi\varphi}=\frac{\frac{1}{v}\left(\frac{2e^2}{v}-\frac{ue}{K}\right)\left({p}^2+\frac{2}{v}\right)
                       +\frac{Ke^3}{v^3}\left(\frac{u}{K^2}-\frac{4}{Ke}+\frac{6u^2v}{Kel^2}\right)}
        {\left({p}^2+\frac{2}{v}\right)\left({p}^2+m^2\right)},\nonumber\\
 && C_{ff}=\frac{\frac{e^2}{v^2}\left(\frac{u}{K^2}-\frac{4e}{Kv}\right)\left({p}^2+\frac{2}{v}\right)^2
           +\frac{e^2}{v^2}B\left({p}^2+\frac{2}{v}\right)+\frac{e^4}{v^4}A}
      {\left({p}^2+\frac{2}{v}\right)^2\left({p}^2+m^2\right)},
\end{eqnarray}
where
\begin{eqnarray}
A&=&\left(\frac{8}{v}-\frac{12u^2}{l^2}\right)\left(\frac{u}{K^2}-\frac{5e}{2Kv}\right)+\frac{e^2}{v^2}
       \left(\frac{u}{K^2}-\frac{4}{Ke}+\frac{6u^2v}{Kel^2}\right),\nonumber\\
B&=&\frac{4e^2}{v^2}\left(\frac{3e}{Kv}-\frac{u}{K^2}\right)
                -\frac{2e^2}{v^2}\left(\frac{u}{K^2}-\frac{4}{Ke}+\frac{6u^2v}{Kel^2}\right).\nonumber
\end{eqnarray}
Finally, similar to the $K=0$ case, under the source condition $T=\frac{e^2}{v^2}J_f$,
we obtain the Fourier-transformed on-shell amplitude induced by the external sources as
\begin{equation}
 \bar{A}^K_{\pm,w}=\frac{u}{8}\int d^2p \frac{1}{{p}^2+m_{\pm,w}^2}\left(\frac{2e}{Kv}J_f-J_\varphi\right)^2.
\end{equation}
As a result, we explicitly observe that a new pole of $1/(p^2+m^2)$ is non-vanishing and
$J_\varphi$-dependent amplitudes appear when comparing with the $K=0$ case.

\end{document}